\documentclass[12pt]{article}
\usepackage{graphicx}
\usepackage{url}

\newcommand\pubnumber{SNSN-323-63}
\newcommand\pubdate{\today}

\def\Title#1{\begin{center} {\Large #1 } \end{center}}
\def\Author#1{\begin{center}{ \sc #1} \end{center}}
\def\Address#1{\begin{center}{ \it #1} \end{center}}

\newcommand\pubblock{\rightline{\begin{tabular}{l} \pubnumber\\
         \pubdate  \end{tabular}}}
\newenvironment{Abstract}{\begin{quotation}  }{\end{quotation}}
\newenvironment{Presented}{\begin{quotation} \begin{center} 
             PRESENTED AT\end{center}\bigskip 
      \begin{center}\begin{large}}{\end{large}\end{center} \end{quotation}}

\def\beq{\begin{equation}}
\def\eeq#1{\label{#1}\end{equation}}
\def\eeqn{\end{equation}}

\def\beqa{\begin{eqnarray}}
\def\eeqa#1{\label{#1}\end{eqnarray}}
\def\eeqan{\end{eqnarray}}

\let\bar=\overbar

\def\Dslash{\not{\hbox{\kern-4pt $D$}}}
\def\dslash{\not{\hbox{\kern-2pt $\del$}}}

\def\msb{{\bar{\ssstyle M \kern -1pt S}}}

\begin{document}
\begin{titlepage}
\pubblock

\vfill

\Title {Statistical Issues in Neutrino Physics Analyses}
\vfill
\Author{Louis Lyons}
\Address{Blackett Lab, Imperial College, London SW7 2BK, UK \\
and\\
Particle Physics, Oxford OX1 3RH, UK}
\vfill
\begin{Abstract}
Various statistical issues relevant to searches for new physics or to parameter determination in analyses of data in neutrino experiments are briefly discussed.
\end{Abstract}
\vfill
\begin{Presented}
NuPhys2015, Prospects in Neutrino Physics \\
Barbican Centre, London UK, December 2015
\end{Presented}
\vfill
\end{titlepage}
\def\thefootnote{\fnsymbol{footnote}}
\setcounter{footnote}{0}


\section{Introduction}
Experiments in High Energy Physics are expensive, both in terms of money and in Physicists' time. In analysing data, it is thus important to invest effort in understanding statistical techniques, which are inexpensive, in order to extract the maximum of information. One must also avoid extracting more information than is justified by the data,

There are several types of statistical procedures that we use. These include: 
\begin{itemize}
\item{Parameter estimation e.g. $\sin^2 2\theta_{23}$. This involves the best value of the parameter and also an estimate of its uncertainty.  Sections \ref{Combining} to
\ref{UL} are examples of this.}
\item{Goodness of Fit. This involves seeing whether a particular data set is consistent with a given hypothesis. e.g. Are results on neutrino mixing consistent with there being just 3 neutrino flavours? See Sections \ref{LEE} to \ref{5sigma} below.}
\item{Hypothesis Testing. The predictions of two (or more) hypotheses are compared with the data, to see which hypothesis is preferred. e.g. Normal or inverted neutrino mass hierarchies. See Sections  \ref{Comparing} to \ref{Example}. }
\end{itemize}
For most of these procedures, there are two fundamental but very different approaches: Bayesian and Frequentist\cite{BF}, as well as ad hoc techniques such as $\chi^2$. A particular implementation of the frequentist approach is the Feldman-Cousins method\cite{FC}, which has many positive features.

In this talk, several different topics of relevance to measurements of parameters or to searches for new phenomena are briefly discussed. 

\section{Combining results}
\label{Combining}
When two or more analyses measure the same physical quantity, together with their uncertainties and correlations, it is possible to combine the results to obtain a best value and its uncertainty. BLUE\cite{BLUE} is a method of doing this. 

A feature of BLUE is that if a counting experiment is performed over two equal time periods, and there are $100\pm10$ and $1\pm1$ counts in the two periods\footnote{It is of course a crime to combine two such discrepant measurements. I have chosen these values merely to give a dramatic numerical result, but a similar effect exists for more consistent measurements.}, the  BLUE combined value is $2\pm1$. This unfortunate result is a consequence of using {\bf estimated}  uncertainties to determine the relative  weights of the measurements, rather than the {\bf `true'} uncertainties. `Iterative BLUE'\cite{LMS} or even better a likelihood approach avoids this problem.

Another feature of BLUE is that for two correlated measurements, the best combined result will lie outside the range of the two measurements if the correlation coefficient 
$\rho$ is larger than $\sigma_1/\sigma_2$, the ratio of the two uncertainties. This in fact is not unreasonable as two very correlated measurements are likely to lie on the same side of the true value. However, if the uncertainties or $\rho$ are incorrectly estimated, the amount of extrapolation required for the best estimate may be incorrect. It was thus  common practice with the LEP experiments that when a parameter was determined by more than one analysis using the same data sample (and hence the estimates were very correlated), rather than trying to combine the results, a single analysis (chosen in advance of obtaining the actual values) was used as the one from which to quote the result, and the remainder were used for confirmation.

When two (or more) estimates are made of a pair of parameters $\alpha$ and $\beta$, the combined uncertainties 
$\sigma_{\alpha}$ and $\sigma_{\beta}$ can be much smaller than the individual ones, if the correlation coefficients $\rho_1$ and $\rho_2$ for the two analyses are very different.  An example from Cosmology is that there are several different methods for estimating $\Omega_{DE}$ and $\Omega_{DM}$, the fractional densities of Dark Energy and Dark Matter in the Universe. Although individual analyses determine $\Omega_{DE}$ poorly, the combination determines it well\cite{DarkE}.    
 
It is important to realise that in general it is far better to combine the data of the individual experiments and to perform a single combined analysis, rather than simply to combine the results.

\section{Coverage}
For a given statistical procedure for estimating the acceptable range for a physical parameter at a particular confidence level, because of statistical fluctuations, these ranges will vary. The fraction of ranges which include the true value of the parameter is known as `Coverage'. Ideally we would like the coverage to be independent of the true value of the parameter, and equal to the chosen confidence level.

An example is estimating a Poisson parameter $\mu$, by making one observation of the number of counts $n$, and using the likelihood method with the 
$\Delta(lnL) = 1/2$ rule for obtaining its range. The coverage for this case is shown in the figure on page 10  of ref. \cite{Heinrich}, and is distinctly non-constant, with regions of $\mu$ where the coverage is significantly below its nominal value. To avoid this undercoverage, a Neyman construction  method can be used for obtaining the confidence intervals.

It is important to realise that coverage is a property of a statistical procedure and does not apply to your actual measurement. 

\section{Blind Analysis}
Blind analysis was first used in Particle Physics when Luis Alvarez noticed that an experiment claiming the discovery of free quarks\cite{Freeq}  involved several corrections to the raw data and also decisions on which experimental runs were to be accepted. The net result of his suggestion to use a blind analysis was that no further results were published by that group on free quarks.

The advantage of performing a blind analysis is that the results cannot be affected by the (subconscious) wish of the experimentalists to produce a particular result. Usually, however, it takes longer to perform a blind analysis. 

There are various ways of performing blind analyses, so if you plan to use this approach, careful thought is needed on how this should be implemented. An aim is to allow you to see as much of the real data as possible, without being able to work out how this affects the result of the analysis. An unwritten rule is that after the analysis is unblinded, nothing should be changed before publication, unless people would regard you as stupid not to do so. e.g. If in a search for dark matter all  observed candidates appeared at midnight between the first Saturday and Sunday of each month. 

If several groups within a Collaboration are analysing the same data with the same physics aim, it should be decided in advance how the Collaboration's result should be presented, rather than looking at the separate numerical results and then making a decision. 

\section{Background systematics}
In almost every analysis, estimating systematic uncertainties involves a lot more effort than statistical uncertainties. Here we describe just one particular example.

Consider an analysis which is designed to estimate the magnitude of a peak on top of a smooth background.  We can perform a likelihood fit, assuming that we know the expected shape of the signal and of the background. The latter may be estimated from theoretical considerations, from Monte Carlo simulation or from control samples, but we are going to consider the case where the actual spectrum is described by a signal shape and a particular functional form with some free parameters for the background e.g.  a third order polynomial. Then the fit to the data will give the signal strength and its statistical uncertainty, which includes the fact that the parameters of the functional form are not precisely known even after the fit. There is however also a systematic uncertainty due to the fact that the chosen functional form may not be appropriate. One method for dealing with this is to see how much the signal strength changes when different functional forms are used to describe the background. But if an alternative functional form results in a best fit which is considerably worse than that for the favoured function, the result is ignored and does not contribute to the systematics.
 
A disadvantage of this approach is that a hard cut is used to decide whether an alternative function does or does not contribute to the systematics. A recent analysis for assessing the strength of the Higgs boson signal uses a new approach\cite{Dauncey}. The log-likelihoods using various functional forms for the background  are determined as functions of the signal strength. Then the envelope of all these different log-likelihoods is used to obtain the best value and range for the signal strength, which now includes the systematic associated with the uncertain functional form for the background. This method is essentially an extension of the traditional profile likelihood method for dealing with a continuous nuisance parameter, to the case of discrete choices for the systematic. 
   
\section{Upper Limits}
\label{UL}
When we look for some new phenomenon (e.g. a $4^{th}$ generation neutrino) and do not observe a signal, it is common to quote an upper limit on the possible signal strength (usually as a function of the mass of the new signal).  Such upper limits can be useful for excluding various proposed models, either completely or over part of their parameter space. The best historic example of this is the Michelson-Morley experiment\cite{MM}, which set an upper limit on the speed of the Earth through the hypothesised aether. This limit was stringent enough to lead to the death of the aether hypothesis. Soon after, Einstein developed Special Relativity.

There are many different methods for setting upper limits. Many of these were discussed extensively at the first two PHYSTAT meetings\cite{CERN_CLW}\cite{FNAL_CLW}.  The various methods can give different limits. This was illustrated in a plot by Narsky\cite{Narsky}, showing the various limits that are obtained in a Poisson counting experiment: limits are set on a possible signal strength $s$ when the expected background is $b$ and there are $n$ observed events. Especially if $n$ is less than $b$, there is a  large  range of upper limits on $s$ given by the various methods. In quoting upper limits, it is thus crucial to be explicit about the method used.

\section{Look Elsewhere Effect}
\label{LEE}
A peak in a mass spectrum could be due to either New Physics (e.g. neutrino-less double beta decay) or to a statistical fluctuation of the background. The probability of having a Poissonian fluctuation at least as large as the one seen at that location in the spectrum, with an assumed background level there, is known as the {\bf local} $p$-value. A very small $p$-value makes a background fluctuation sound unlikely, and hence lends credence to the New Physics hypothesis.

However the local $p$-value can understate the probability of a fluctuation, as a peak elsewhere in the analysis (e.g. at a different location in the mass spectrum) could also be exciting. So we really need the {\bf global} $p$-value, which is the probability of having a fluctuation at least as significant as the one we observed anywhere relevant.

The trouble is that `anywhere' is poorly defined. For example, it can depend on whether you are a graduate student, concerned only with your own analysis, or the Director General of CERN, who may well be worried about claims by any CERN experiment which may turn out to be just a statistical fluctuation.

In the event of a discovery claim, the informal consensus is to publish your local $p$-value, and a global one too. The latter could relate to a statistical fluctuation anywhere relevant in the spectrum in which the possible signature of new physics was seen. In any case, it is important to state explicitly what you are assuming in calculating your global $p$-value.

\section{Meaning of $p$-values}
A $p$-value is the probability under an assumed hypothesis of obtaining a result at least as extreme as ours. Depending on the possible alternative hypothesis, this might involve a one-sided or a 2-tail definition of `more extreme'.

It is crucial to realise that this is not the same as the probability of the assumed hypothesis being true, given the observed data. First, the probability of a hypothesis being true is not a frequentist concept, as frequentist probability requires the possibility of a large number of essentially identical trials, while a hypothesis is generally simply true or false. Secondly it incorrectly interchanges 
$P(A|B)$ and $P(B|A)$. Thus statements such as `Given this data, we are 99$\%$  sure that the $\tau$-neutrino  exists' are simply not correct.

Sometimes $p$-values are criticised because they are misunderstood; and because they tend to be smaller than likelihood ratios and hence to overstate the evidence against the null hypothesis. Concerning the former, it simply needs people to be better educated as to what $p$-values are. There are no suggestions that relativity should be ignored, because people don't understand it. The comparison with likelihood ratios is misleading, because they are different quantities; $p$-values refer just to one hypothesis, while likelihood ratios involve two hypotheses. It is like trying to decide whether the length of a mouse in light years  or its mass relative to that of the Earth is better for describing the size of the mouse.    

\section{$P(A|B) \ne P(B|A)$}
My grand-daughter's example is\cite{Kezia}: $\\$
The probability of a person eating bread, given the fact that they are a murderer, is greater than 99$\%. \\$
The probability that a person is a murderer, given that they eat bread, is about one in a million. 

\section{Significance}
When a bump is observed in a mass spectrum, its significance is defined by converting its $p$-value into the number of standard deviations away from the centre of a Gaussian distribution that is required in order for the one-sided tail area to equal $p$ \footnote{The only reason for doing this is that the number of standard deviations is a number that is easier to remember than a $p$-value. e.g. 5 standard deviations corresponds to $p = 3\ 10^{-7}$.}.

In dealing with peaks in mass distributions,  people often try to use the  approximation $S/\sqrt B$ (or some variant of this) for the significance; here $S$ is the number of possible signal events in the observed spectrum above the estimated background $B$ in the relevant region. This is based on assuming that the Poisson distribution can be approximated by a Gaussian. This can be a poor approximation, because:
\begin{itemize}
\item{The `Poisson is like a Gaussian' approximation is better in the central region of the distributions than in the tails. It is the upper tail which is relevant here.}
\item{No account is taken of any possible LEE factor in reducing the significance.}
\item{Any uncertainty in $B$ is ignored.}
\end{itemize}

Another situation in which the approximation can be bad is in designing an analysis to maximise the expected $S/\sqrt B$. It could be that very hard cuts reduce the background $B$ to $10^{-4}$ events, while the expected signal $S$ is reduced to $10^{-2}$ events. Then $S/\sqrt B$ is 10, which sounds good. But the expected signal is such that $99\%$ of the time there would be no observed events, which is not ideal in a search for a signal.

\section{Combining $p$-values}
Just as it is possible to combine results of different measurements of the same parameter, in searches for disagreements with the null hypothesis of no new physics, the $p_i$-values for different independent searches can be combined. However, it should remembered that
\begin{itemize}
\item{The procedure is not unique. As Cousins has pointed out, the combined $p$-value should be uniformly distributed in the range 0 to 1, and is constructed from the individual $p_i$-values, whose joint distribution should be uniform over a hypercube. Clearly there are many ways of achieving this transformation. Which is best depends on more details about the individual analyses than just their $p$-values. }
\item{Different experiments can be inconsistent with the null hypothesis but in different ways.}
\item{As with combining measurements, a single analysis using the data from all the separate experiments is better than just combining their $p$-values.}  
\end{itemize}

Despite all this, if it is desired to go ahead and combine the $n$ individual independent $p_i$-values, one recipe is to calculate the probability $P_{comb}$ that the product of the $p_i$ is smaller than the observed value $z = \Pi  p_i$. For uniformly distributed $p_i$, this is given by 
\begin{equation}
P_{comb} = z\Sigma (-ln z)^k , 
\end{equation}
where the summation over $k$ runs from zero to $n-1$.

\section{Why $5\sigma$ for discovery?}
\label{5sigma}
One of the standard requirements for claiming a discovery in particle physics is that the $p$-value for the null hypothesis of no new physics is below $3\ 10^{-7}$. Statisticians pull fun of us, saying that no probability distribution is known with any accuracy so far into its tails.

The justification used in High Energy Physics for such a stringent requirement includes:
\begin{itemize}
\item{Past History. There are plenty examples of $3\sigma$ and $4\sigma$ effects that have gone way when more data are collected.}
\item{Under-estimated systematics. If a measurement is dominated by systematics, which have in fact been under-estimated by a factor of 2, what appears to be an interesting $4\sigma$ effect is in fact a much more mundane $2\sigma$ observation.}
\item{Look Elsewhere Effect. See Section \ref{LEE}}
\item{Subconscious Bayes probabilities. Even though a frequentist  approach may have been used for Hypothesis Testing, we may want to know the relative probabilities $P(H_i$|data) of the null and alternative hypotheses being true.  These are given by Bayes' Theorem as
\begin{equation} 
\frac{P(H_1|data)}{P(H_0|data)} = \frac{P(data|H_1)}{P(data|H_0)} \frac{\pi_1}{\pi_0}
\end{equation}
Here the first term on the right hand side is the likelihood ratio, but this has to be multiplied by the ratio of the prior probabilities 
$\pi_i$ of the hypotheses. If 
$H_1$ is very speculative (e.g. energy is not conserved), $\pi_1$ would be very small compared with the energy conservation prior $\pi_0$ and thus the likelihood ratio would have to be enormous in order for us to believe (and be prepared to publish) that energy is not conserved. 

This is a numerical formulation of the aphorism `Extraordinary claims require extraordinary evidence.'}
\end{itemize}

It is clear that the importance of the last three items varies very much from experiment to experiment, and so it does not seem reasonable to have the same $5\sigma$ requirement for all analyses. Ref. \cite{LL_5sigma} is an  attempt to provoke discussion on this issue. 

\section{Comparing 2 Hypotheses}
\label{Comparing}
There are two different situations in which we may want to compare our data with two (or perhaps more) hypotheses:
\begin{itemize}
\item{We have a number of tracks or events, and are trying to assign them to different hypotheses e.g. jets are assigned as being produced by $b$-quarks or light quarks; which type of particle initiated Cerenkov rings; etc. }
\item{The hypotheses refer to the conclusion of the experiment e.g. Do we have evidence for neutrino-less double beta decay? Is the normal or inverted neutrino mass hierarchy preferred?}
\end{itemize} 

The desiderata for the two cases are not necessarily the same. For the former case, we essentially have to make a decision each time we are presented with a track. Errors of the First Kind are when we fail to accept the true hypothesis. This corresponds to a loss of efficiency, and should occur at a well-defined rate. Errors of the Second Kind are when we accept a hypothesis when a different hypothesis is true; it results in a reduction in purity of our selected sample, and is usually harder to calculate. These effects are to be taken into account in the analysis. 

For the result of the experiment, it is possible not to make a decision among competing hypotheses. Here getting the wrong answer is more unfortunate than in the event selection scenario.

In both situations, multivariate techniques are often used to provide a variable on which to base one's choice. The probability density functions for the  two hypotheses are obtained (often by Monte Carlo simulation), and form the basis of choosing between the hypotheses. There are many ways of doing this; these include
\begin{itemize}
\item{The log-likelihood ratio}
\item{Difference in $\chi^2$}
\item{$p$-values of test statistic}
\item{Bayesian methods e.g. posterior odds, Bayes factor, Akike information criterion, etc.}
\item{Minimum cost}.
\end{itemize}
For more detail,  see, for example, ref. \cite{Comp_Hyps}.

\section{Wilks' Theorem}
\label{Wilks}
This applies to seeing which of two hypotheses provides a better description of some data. It involves the difference $\Delta S$ in the weighted sums of squares $S_0$ and $S_1$ for the two hypotheses (or almost equivalently the log-likelihood ratio). If the correct hypothesis is $H_0$, then $S_1$ should not be much smaller than $S_0$. Wilks' Theorem quantifies this by stating that assuming the conditions below are satisfied, $\Delta S = S_0 - S_1$ should be distributed as $\chi^2$ with the  number of degrees of freedom equal to the difference in the number of free parameters for the two hypotheses. 

The conditions are
\begin{itemize}
\item{The two hypotheses must be nested. That is, it must be possible to reduce $H_1$ to $H_0$ by suitable choice of $H_1$'s extra parameters. For example, a straight line and a quadratic expression are nested, while the alternative neutrino mass hierarchies are not.}
\item{In order to reduce the larger hypothesis $H_1$ to the smaller one, all the extra parameters of $H_1$ must all be well-defined and not be on their physical boundaries.} 
\item{The data must be asymptotic.}
\end{itemize}

If these conditions are not satisfied, all is not lost as $\Delta S$ can still be used to choose between the hypotheses, but its distribution must be determined, for example by simulation.  

\section{Example of Hypothesis Testing}
\label{Example}
One of the current big issues in neutrino physics is whether the neutrino mass hierarchy is normal or inverted. This is almost an example of the classical `Simple versus Simple' Hypothesis Testing\footnote{A simple hypothesis is one which is completely specified, without any arbitrary parameters.  In contrast, a composite hypothesis does have nuisance parameters, e.g. those  related to experimental features such as jet energy scale.}. The only nuisance parameter here is the phase $\delta$ of the possible $CP$ violation. This is best dealt with by presenting results as a function of $\delta$, rather than integrating over it.

As outlined in Section \ref{Comparing}, the procedure involves producing the expected probability density functions of the data statistic for the two hypotheses, and comparing with the actual data statistic. This is very similar to the procedure for deciding whether the data on Higgs decays at the Large Hadron Collider are consistent with spin-parity of $0^+$ or $0^-$\cite{ATLASH}\cite{CMSH}.   

\section{Conclusions}
There are many resources available. These include 
\begin{itemize}
\item{Textbooks written by experimental Particle Physicists\cite{books}}
\item{Sections on Probability, Statistics and Monte Carlo in the Particle Data Book\cite{pdg}}
\item{The web-pages of the Statistics Committees of large experiments e.g CDF\cite{CDF}, CMS\cite{CMS}, etc.}
\item{Software packages e.g RooStats\cite{RooStats}. This can be unfortunately easy to implement, with the result that it can be used by people who don't really understand the subtleties of what they are doing.} 
\end{itemize} 

If you have a statistics problem in your analysis which you think is new, before attempting to devise your own `Reinventing the wheel' method of solving the problem, it is worth looking to see whether other Particle Physicists or Statisticians have already done so. It is in general better to use a Statistician's circular wheel than your own hexagonal or square one.
\vspace{0.2in}


\vspace{0.1in}


\vspace{0.2in}



%

%
\end{document}